\documentclass[apj]{emulateapj}

\usepackage{apjfonts}

\newcommand{\ergs}{ergs s$^{-1}$}
\newcommand{\flux}{ergs cm$^{-2}$ s$^{-1}$}
\newcommand{\intens}{ergs cm$^{-2}$ s$^{-1}$ deg$^{-2}$}

\newcommand{\hst}{{\it HST}}

\newcommand{\chandra}{{\it Chandra}}
\newcommand{\spitzer}{{\it Spitzer}}

\newcommand{\logn}{$\log{N}-\log{S}$}

\newcommand{\im}{\item}

\newcommand{\average}[1]{\ensuremath{\langle#1\rangle} }

\newcommand{\z}{$z_{850}$}
\newcommand{\B}{$B_{435}$}
\newcommand{\V}{$V_{606}$}
\newcommand{\iband}{$i_{775}$}

\newcommand{\hmx}{\citetalias{hick06a}}
\newcommand{\hmxp}{\citepalias{hick06a}}

\newcommand{\hmn}{\citetalias{hick07b}}

\newcommand{\alex}{\citetalias{alex03}}

\newcommand{\baue}{\citetalias{baue04}}

\begin{document}

\title{Can Chandra resolve the remaining cosmic
  X-ray background?}
\shorttitle{REMAINING COSMIC X-RAY BACKGROUND}
\shortauthors{HICKOX \& MARKEVITCH}
\author{Ryan C. Hickox}
\author{Maxim Markevitch\altaffilmark{1}}
\affil{Harvard-Smithsonian Center for Astrophysics, 60 Garden Street,
 Cambridge, MA 02138}
\altaffiltext{1}{Also at Space Research Institute, Russian Academy of 
Sciences, Profsoyuznaya 84/32, Moscow 117997, Russia}

\slugcomment{Accepted for publication in The Astrophysical Journal}

\setcounter{footnote}{1}

\begin{abstract}
The deepest extragalactic X-ray observation, the 2 Ms Chandra Deep
Field North (CDF-N), resolves $\sim$80\% of the total extragalactic
cosmic X-ray background (CXB) in the 1--2 keV band.  Recent work has
shown that 70\% of the remaining CXB flux is associated with sources
detected by the {\it Hubble Space Telescope} (\hst).  This paper uses
the existing CDF-N data to constrain the X-ray flux distribution of
these X-ray-undetected \hst\ sources by comparing the number of
0.5--2 keV X-ray counts at the \hst\ positions to those expected for
model flux distributions.  In the simple case where all the undetected
\hst\ X-ray sources have the same 0.5--2 keV flux, the data are best
fit by 1.5--3 counts per source in 2 Ms, compared to a
detection limit (at 10\% completeness) of 9 counts.  Assuming a more
realistic power-law \logn\ distribution [$N(>S)\propto S^{-\alpha}$],
the data favor a relatively steep flux distribution, with
$\alpha=1.1^{+0.5}_{-0.3}$ (limits are 99\% confidence).
This slope is very similar to that previously found for faint normal
and starburst galaxies in the CDF-N.  These results suggest deeper
\chandra\ observations will detect a new population of faint X-ray
sources, but extremely deep exposures are needed to resolve the
remainder of the soft CXB.  In the most optimistic scenario, when the
\hst\ sources have the flattest allowed flux distribution and all the
sources without \hst\ counterparts are detected, observations 5 times
more sensitive than the existing ones would resolve at most $\sim$60\%
of the remaining soft CXB.

\end{abstract}

\section{Introduction}
\label{intro}
Deep X-ray observations, in particular the 1 and 2 Ms Chandra Deep
Fields North and South (CDF-N and CDF-S), have resolved most of the
extragalactic cosmic X-ray background (CXB) at energies $E<4$ keV into
discrete X-ray sources \citep[see][for a review]{bran05}.  The
spectrum of the still-unresolved CXB in the CDFs was directly measured
by \citet[hereafter HM06]{hick06a}, who found that 23\%$\pm$3\% of the
CXB in the 1--2 keV band remained unresolved.  There is also an
unresolved component at energies $E>2$ keV, but the uncertainties are
larger.  In the 1--2 keV band, most of the unresolved flux can be
accounted for by galaxies that are detected in deep {\it Hubble Space
Telescope} (\hst) observations, but are too faint to be detected as
individual X-ray sources.  A stacking analysis by \citet{wors06}
showed a significant contribution from X-ray-undetected \hst\ sources.
\citet[hereafter HM07]{hick07b} measured the remaining CXB spectrum
after excluding all sources detected by \hst\ or by the Infrared Array
Camera (IRAC) on the {\it Spitzer Space Telescope}, and found that
only $7\%\pm3\%$ of the total 1--2 keV CXB remained.

\defcitealias{hick06a}{HM06}
\defcitealias{hick07b}{HM07}
\defcitealias{alex03}{A03}
\defcitealias{baue04}{B04}

There has been significant interest in using deeper \chandra\
observations to resolve even more of the CXB, especially since longer
\chandra\ exposures might probe a new population of sources.  A
simple extrapolation of the total observed \logn\ distribution [which,
fitted by a power law of the form $N(>S)\propto S^{-\alpha}$, has
$\alpha\simeq 0.7$] falls far short of accounting for the entire 1--2
keV CXB.
If the remaining signal is due to discrete sources, an upturn in the
\logn\ at low fluxes is required \hmxp.  Evidence for  a possible upturn in
the \logn\ distribution at $S_{0.5-2\; {\rm keV}}\lesssim 10^{-17}$
\ergs\ was found through a fluctuation analysis using
the first 1 Ms of CDF-N data \citep{miya02}.

These faint objects, with fluxes less than the faintest CDF-N sources
($S_{0.5-2\; {\rm keV}}=2.4\times10^{-17}$ \flux) are likely starburst
and normal galaxies, unlike the majority of the detected X-ray sources
in the CDFs, of which $\gtrsim$75\% are active galactic nuclei (AGNs).
\citet[hereafter B04]{baue04} divided the CDF X-ray sources into
galaxy and AGN subsets, based on their X-ray and optical properties,
and produced separate \logn\ fits to each using power law models.  AGNs
have a shallow \logn\ slope with $\alpha \simeq 0.6$, while the
galaxies have a much steeper distribution with $\alpha\simeq 1.3$.
The galaxies may dominate the source counts at fluxes $\lesssim
10^{-17}$ \ergs, and could produce the upturn in the \logn\
distribution at low fluxes.

The goal of this paper is to constrain the distribution of
X-ray fluxes for the \hst\ sources that are  cumulatively
responsible for most of the unresolved 1--2 keV CXB.  This
study puts limits on the \logn\ distribution only for objects with \hst\
counterparts, not all unresolved X-ray sources (which would
require a fluctuation analysis without the advantage of positional
information).  However, since X-ray-undetected \hst\ sources account
for $\sim$70\% of the unresolved 1--2 keV CXB (or $\sim$16\% of the
total CXB), this analysis provides useful limits on the properties of
X-ray sources fainter than the current CDF-N limit and puts
constraints on what might be observed with longer \chandra\ exposures.

\section{Observations}
\label{data}
The CDF-N consists of 20 separate \chandra\ ACIS-I observations with a
total exposure time of $\simeq$2 Ms \citep[][hereafter A03]{bran01a,
alex03}.  For this analysis, we utilize the coadded image of the CDF-N
in the 0.5--2 keV that was used for source detection by \alex\ (see
their Fig.~3). The ACIS pixel scale is 0.492\arcsec, and the effective
aimpoint (the exposure weighted centroid of the 20 slightly offset
observations) is ($\alpha$, $\delta$, J2000) 12:36:45.9, +62:13:58.
We also use the 0.5--2 keV exposure map shown in Fig.~5 of
\alex.\footnotemark\  The image has a total on-axis exposure time of
1.94 Ms, although the effective exposures at certain positions within
our region of interest are as low as 1.2 Ms because of the chip gaps.
The on-axis exposure is $\sim$2 times longer than that used in \hmn,
in which we filtered the data extensively for background flares (see
\S\ \ref{meas}).  

The nominal flux limit of the \alex\ catalog is
$\approx2.4\times10^{-17}$ \ergs, or $\approx9$ counts in 1.94 Ms.
This corresponds roughly to the flux of the faintest detected sources,
for which the source detection is highly incomplete.  The completeness
of the CDF-N source detection depends on measured source counts, as
shown in Fig.~2~(a) of \baue.  Within 2\arcmin\ from the aimpoint, the
completeness is near zero for $<8$ observed counts, and rises from
$\simeq$10\% for 10 counts to $\simeq$90\% for 20 counts.  We use
this completeness curve in our modeling of the undetected X-ray sources.

\footnotetext{For CDF-N images and exposure maps from
  \alex, see http://www.astro.psu.edu/users/niel/hdf/hdf-chandra.html}

The \hst\ catalog of sources is the same as we used in \hmn\ and comes
from the Great Observatories Origins Deep Survey
\citep[GOODS;][]{dick03}.  Data were taken using the Advanced Camera
for Surveys (ACS) on \hst\ in the \B, \V, \iband, and \z\ bands
\citep{giav04}.  We use the public catalog\footnotemark\ for the \z\
band (sampling wavelengths $\sim$8300--9500 \AA), which has an
approximate magnitude limit of $z_{850} \simeq 27$ (AB) and nominal
positional uncertainty of $\approx$0.1\arcsec.  \hmn\ also excluded
\spitzer\ IRAC sources, but 90\% of these also had \hst\ counterparts,
so we do not consider them here.

We slightly shift the positions of the \hst\ sources to best match the
positions of the brightest 100 \alex\ X-ray sources.  This small offset of
$-0.11$\arcsec\ in $\alpha$ and $-0.28$\arcsec\ in $\delta$ serves to
register the coordinate frames to  $\simeq$0.1\arcsec\
within 2\arcmin\ of the X-ray aim point.  The positional uncertainties of the
\hst\ sources are significantly smaller than the ACIS pixel scale and
so will not affect our results.

\footnotetext{Available at http://archive.stsci.edu/pub/hlsp/goods/catalog\_r1/h\_r1.1z\_readme.html}

\section{Analysis}
\label{analysis}

 \citet{wors06} and \hmn\ showed that \hst\ sources that do not have
X-ray detections in the \alex\ catalog nonetheless contribute
significant flux to the CXB.  Although we cannot measure X-ray fluxes
for these sources individually, the {\em distribution} of observed
counts at the source positions can provide a useful constraint on
their underlying X-ray flux distribution.

\subsection{Measurement of the observed counts distribution}
\label{meas}

\begin{figure}[t]
\epsscale{1.2}
\plotone{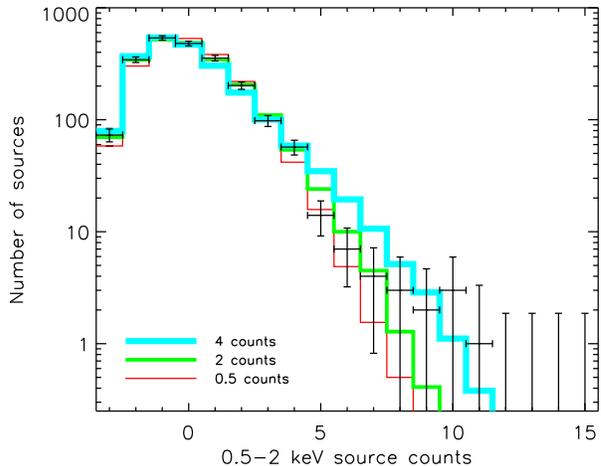}
\caption{Histogram of background-subtracted 0.5--2 keV source counts
  inside the 2184 \hst\ source regions (black crosses).  Solid lines
  show predictions for models with ``delta-function'' flux
  distribution, consisting only of sources with fluxes of 0.5, 2, and
  4 counts, respectively, per 2 Ms.
  \label{figcounts}}
\vskip0.3cm
\end{figure}

To measure this counts distribution, we extract 0.5--2 keV photons at
the \hst\ source positions in circular regions of radius $r_{90}$,
which is an approximation of the 90\% point-spread function (PSF) energy enclosed radius and
 varies as\footnotemark:
\begin{equation}
r_{90}=1\arcsec+10\arcsec(\theta/10\arcmin)^2,
\label{eqnrad}
\end{equation}
where $\theta$ is the off-axis angle.  Because $r_{90}$ increases with
  $\theta$, at large angles many of the \hst\ source regions begin to
  overlap.  Within a 1\arcmin\ radius of the aimpoint, the regions
  overlap over only 9\% of their area, while within a 4\arcmin\
  radius, 28\% of the area overlaps.  To include sufficient numbers of
  sources but still have minimal overlap, we confine our analysis to a
  small region of radius 2.2\arcmin\ around the aimpoint, for which
  $r_{90}$ is between 2 and 3 ACIS pixels ($1\arcsec<r_{90}<1\farcs5$)
  and overlap is only 12\%.  Using a model of the \chandra\ PSF as
  described in \S~\ref{models}, we find that for sources within
  the 2.2\arcmin\ radius, the flux scattered outside the $r_{90}$
  circles is 9.9\%, accounting for the source overlap.  In order to
  exclude the flux from detected X-ray sources, we only consider those
  \hst\ sources that are outside large exclusion regions of $4.5r_{90}$--$9
  r_{90}$ around the \alex\ sources, which exclude $>99.9\%$ of the
  flux from these sources (see \S\ 3.2 of \hmx).  We also exclude one
  extended source from the catalog of \citet{baue02}, with an
  exclusion region of radius 0.75\arcmin, or 1.5 times the measured
  source extent.  This leaves $N_{HST}=2184$ \hst\ sources over an
  area of 12.2 arcmin$^2$.

 \footnotetext{\chandra\ Proposer's Observatory Guide, http://cxc.harvard.edu/proposer/POG/}

We focus our analysis on the 0.5--2 keV band, as opposed to the 1--2
keV band as in \hmn, for two reasons.  First,
it allows us to use the published CDF-N image from \alex, exactly the
same image used for source detection.  Second, it
enables straightforward comparisons between our results and previous
measurements of the soft X-ray \logn, which usually are measured for
0.5--2 keV.  One  drawback of the 0.5--2 keV band is that it
also includes significant diffuse signal at $E<1$ keV from local and
Galactic emission \hmxp; however, this diffuse component can be
accurately subtracted from the fluxes measured in the small \hst\
source regions (see below), and so does not significantly affect the
measured counts distribution.  We assume that the fraction of the
extragalactic CXB resolved by \hst\ sources is the same in the 0.5--2
and 1--2 keV bands, since faint X-ray sources tend to have power law
spectra with photon index $\Gamma \sim 1.4$, similar to the total
extragalactic CXB \citep[e.g.,][see Fig.~13 of \hmx]{rosa02}.

A key element in this analysis is the background subtraction, since
the background not associated with the \hst\ sources is $\sim$4 times
larger than the source flux within the small source regions.  Our
background-subtraction procedure is different from that used in \hmx\
and \hmn.  Here the background consists of (1) local and Galactic
diffuse sky emission, (2) flux from faint unresolved X-ray sources
that do not have \hst\ counterparts and whose positions we assume are
random in the field, and (3) the instrumental background, with both
quiescent and flaring components (see \S~4 of \hmx).  The full 1.94 Ms
image from \alex\ includes some flux from background flares, which can
be eliminated using detailed light-curve analysis, as we did in \hmx\
and \hmn.  Here, in order to maximize the available exposure time we
do not clean for flares.  The flaring background is distributed nearly evenly
across the X-ray image, so we can treat it as a component of the total
diffuse background.

\begin{figure}[t]
\epsscale{1.1}
\plotone{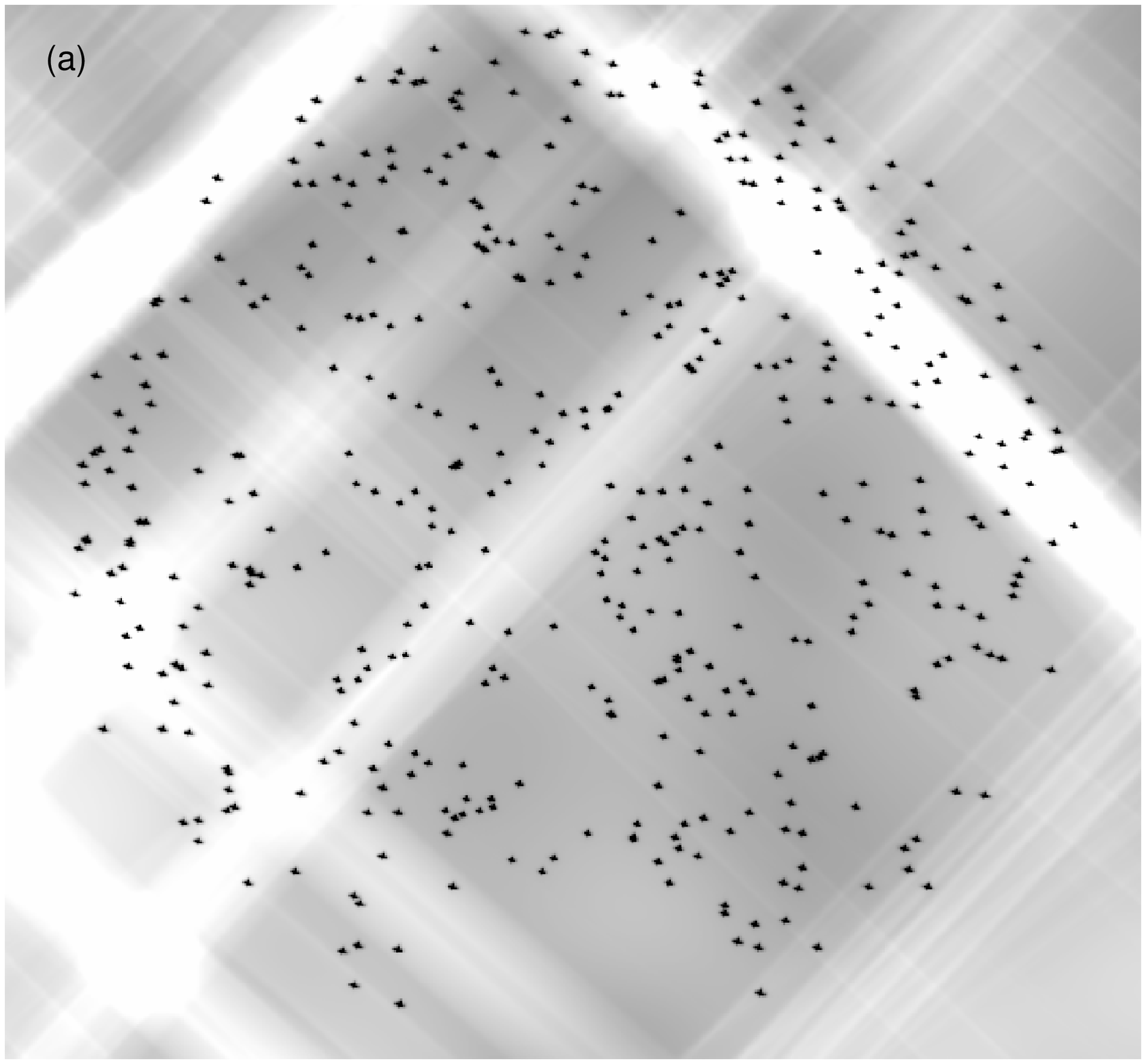}
\plotone{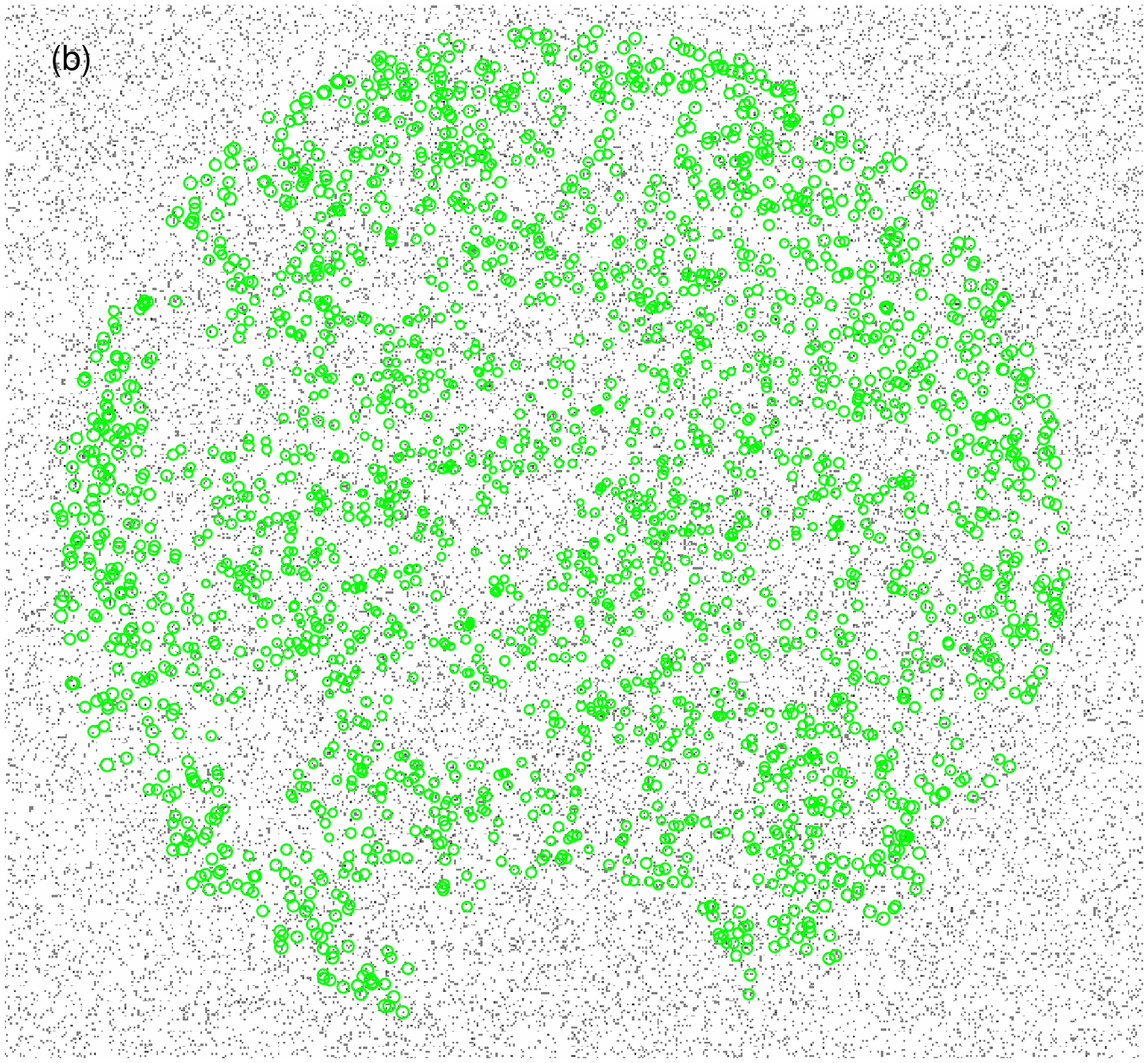}
\caption{(a) Simulated photon probability map and (b) corresponding
  sky image, for a model in which all \hst\ sources with X-ray
  counterparts have average fluxes of 2 counts per 2 Ms exposure.
  The extraction regions of radius $r_{90}$ around all X-ray sources are shown in (b).  The large empty regions within the area of radius 2.2\arcmin\ are
  the excluded regions around detected X-ray point and extended sources. \label{figmap} }
\vskip0.3cm
\end{figure}

All the above components of the background are distributed relatively
smoothly across the field of view, so we can accurately estimate the
intensity and spatial distribution of the background by performing a
wavelet decomposition \citep{vikh98b} on the 0.5--2 keV image itself.
The small-scale wavelet components include all detectable
X-ray sources, while the largest scale (40\arcsec) component
approximates the background.  This component also includes signal from
the X-ray-undetected \hst\ objects, whose fluxes we want to analyze.
To remove this signal, we measure the number of 0.5--2 keV
counts outside the \hst\ and X-ray exclusion regions but inside the
2.2\arcmin\ circle, and then normalize the large-scale wavelet image
to match the surface brightness in this area. We approximately account for the
fact that 9.9\% of the flux from the \hst\ sources is scattered outside
the $r_{90}$ source regions.  This normalization is given by
\begin{equation}
N_{B}=\frac{I_B-0.099(I_{HST}-M_{HST} I_B / M_B)}{M_B},
\label{eqnbgnorm}
\end{equation}
where $I_B$ and $M_B$ are the number of counts in the regions outside
the \hst\ and X-ray source regions and $I_{HST}$ and $M_{HST}$ are the total counts
inside the \hst\ regions, in the image ($I$) and unscaled wavelet
background model ($M$), respectively.
For the \alex\ 0.5--2 keV image, $I_B=14,691$, $M_B=15,584$, $I_{HST}=4806$, and
$M_{HST}=4175$ counts, resulting in $N_{B}=0.937\pm0.08$ ($1 \sigma$
statistical error).

Subtracting this background leaves $I_{HST}-N_{B}M_{HST}=894$
 net counts associated with the \hst\ sources.  Using the average
 counts-to-flux conversion from \alex\ for sources within 2.2\arcmin\
 of the aimpoint, this corresponds to a 0.5--2 keV surface brightness
 of $\sim$$8\times10^{-13}$ \intens\ over the 12.2 arcmin$^2$ area.
 Assuming a $\Gamma=1.4$ power-law spectrum, this agrees well with the
 1--2 keV brightness of $5.7\times10^{-13}$ \intens\ for the \hst\
 sources from \hmn, derived from a completely different procedure.  

To derive the distribution of source counts, we measure the counts in
 each of the 2184 \hst\ source regions after subtraction of the
 ``diffuse'' background model and produce the distribution shown in
 Fig.~\ref{figcounts}.  We include bins up to 15 source counts, which
 corresponds to the $\sim$70\% completeness limit for the source
 detection (\baue). Although the histogram bins with $\geq$13 counts
 contain no sources, we include them because they help constrain some
 model flux distributions; the exact upper bound does not strongly
 affect our results.

Because many \hst\ regions overlap, the total number of counts in the
 histogram is 1048, larger than the total of 894 counts mentioned
 above.  We also determine the average exposure time for each \hst\
 region, which varies from $\sim$1.2 Ms in the chip gaps to 1.94 Ms on
 axis, with a mean value of $\average{t_{\rm exp}}\simeq1.8$ Ms; this
 is used below.

\subsection{Comparison to \logn\ models}
\label{models}
To constrain the X-ray flux distribution of the undetected \hst\
sources, we now compare the observed counts histogram to the
predictions for various model \logn\ curves.  Poisson fluctuations
cause the observed {\em counts} distribution to differ from
the underlying {\em flux} distribution.  In addition, we cannot simply evaluate
predictions of the \logn\ models analytically, because many source
regions overlap and share the same photons, so 
we must account for the spatial distribution of the \hst\ sources and
the variation in the \chandra\ PSF.
Therefore, we produce simulated X-ray images
for each model \logn\ distribution  and then derive count histograms that we
can directly compare to the CDF-N histogram.  For each model \logn, we use the
following procedure:

\begin{enumerate}

\begin{figure}[t]
\epsscale{1.1}
\plotone{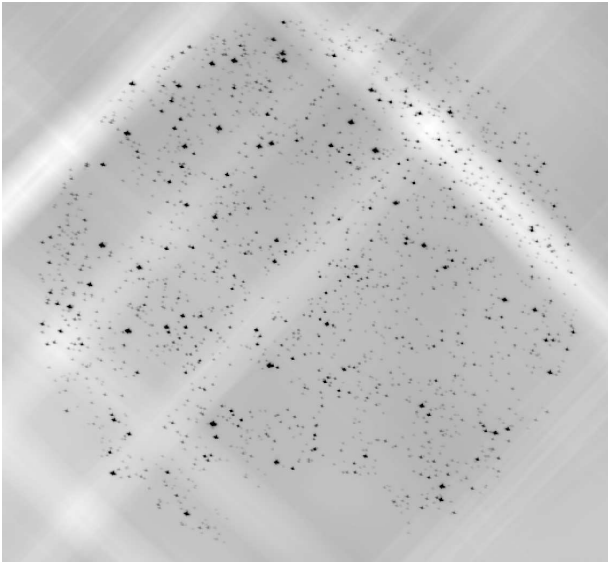}
\caption{ Probability
  map similar to Fig.~\ref{figmap}(a), but for a power-law \logn\
  model with
  $\alpha=1$.
  \label{figmap2}}
\vskip0.5cm
\end{figure}

\im We produce a model flux distribution in units of 0.5--2 keV counts
in 2 Ms (2 Ms rather than 1.94 or 1.8 Ms is used for
simplicity). The fluxes $S$ range from 0.05 to 30 counts per 2 Ms
(which extends sufficiently above the source detection limit) in
$n_{\rm flux}=600$ bins 0.05 counts wide.  Our final results are not
highly sensitive to the flux bounds or bin size.  We define $N_{\rm
src}(S)$ to be the number of sources in each bin of flux $S$ in
the 12.2 arcmin$^2$ area.

\im We normalize the flux distribution to produce, on average, the
observed 894 total source photons in the \hst\ regions.  The $1
\sigma$ statistical uncertainty in this value is $\pm80$ counts,
mostly owing to fluctuations in the total counts before subtraction of
background, but also includes uncertainty in the background
normalization (Eqn.~\ref{eqnbgnorm}), which corresponds to $\pm31$
counts.  The error on the total number of source counts corresponds
directly to uncertainty in the normalization of the unresolved \logn\
distribution.  However, as we show in \S~\ref{simulating}, statistical
fluctuations do not affect our constraints on the {\em shape}
of the \logn\ distribution, so for technical simplicity we ignore
this uncertainty when constraining the \logn\ slope.

In normalizing the model \logn\ distribution, we account for the
scattering of flux outside the $r_{90}$ regions, as well as the
probability that a source of a given flux would be detected in the
CDF-N and not associated with the ``unresolved'' \hst\ sources.  For a
given flux $S$, the average number of observed source counts inside
the source regions is
\begin{equation}
S_{90}(S)=0.9\frac{\average{t_{\rm exp}}}{\rm 2 \;
  Ms}S.
\label{eqnsninety}
\end{equation}
 The average total number of counts, including
background, in each $r_{90}$ circle is 
\begin{equation}
S_{\rm tot}(S)=S_{90}(S)+\average{S_B},
\label{eqnstot}
\end{equation}
where $\average{S_B}=2.03$ is the mean number of background counts per source region.  For the purposes of normalizing the
model \logn\ only, we assume that the exposure time and number of
background counts in all the \hst\ source regions are equal to
$\average{t_{\rm exp}}$ and $\average{S_B}$, although they vary by up
to $30$\%.  As we show below, this gives a sufficiently accurate
normalization to match the number of observed counts.

 In the observed image, we have excluded detected X-ray sources using
much larger regions of radii $>4.5 r_{90}$ to avoid PSF-scattered
flux.  To make the models consistent with this exclusion, we must remove
sources from the simulated image that would be ``detected'' in the
\alex\ catalog.  For a given source flux, there is a probability
(given by Poisson scatter and source detection completeness) that a
source will be detected and thus removed from the image; we estimate
this probability and include it in the normalization as follows.

For a given $S$ and $\average{S_B}$, the number of observed counts is
determined by a Poisson distribution with mean $S_{\rm tot}$.  For $j$
total counts, the measured source counts will be $S_j\equiv
j-\average{S_B}$.  Here we define $f_c(S_j)$ to be the
probability of detecting a source with $S_j$ source counts within an
$r_{90}$ circle.  We estimate $f_c$ from the \baue\ completeness curve (which is
for aperture-corrected total counts), by multiplying their $x$-axis by
0.9.

On average, the number of counts contributed by a source with flux $S$
to the total source counts in the $r_{90}$ regions around {\em
undetected} sources is $S_{90}f_{\rm nondet}$, where
\begin{equation}
\displaystyle f_{\rm nondet}(S,\average{S_{B}})=
\frac{\displaystyle \sum_{j=0}^{\infty} S_j P_P(j,S_{\rm
    tot})[1-f_c(S_j)]}{\displaystyle \sum_{j=0}^{\infty} S_j P_P(j,S_{\rm tot})},
\label{eqnnodetect}
\end{equation}
and where $P_P(j,S_{\rm tot})$ is the Poisson
probability of observing $j$ counts for a mean  of
$S_{\rm tot}$.

On average, the total number of background-subtracted source counts in
all the $r_{90}$ regions associated with X-ray-undetected \hst\
sources is therefore
\begin{equation}
\displaystyle C_{\rm src}=\sum_{i=1}^{n_{\rm flux}}  N_{\rm src}(S_i) S_{90}(S_i) f_{\rm
  nondet}(S_i,\average{S_{B}})
.
\label{eqnnorm}
\end{equation}
We  normalize the model flux distribution $N_{\rm src}(S)$ so
that $C_{\rm src}=894$.

We stress that this normalization is not a free parameter that we fit
to the observed histogram.  Instead, we normalize the model \logn\
{\em before} producing the simulated images, so that on average, they
match the total number of source photons observed in the sky image.
This normalization approach proves sufficient; the resulting simulated
images (that properly include variations in the exposure, background,
and PSF, as described in the following section) produce, on average,
886 counts from ``undetected'' \hst\ sources, within 1\% of the target
value of $C_{\rm src}=894$.  We stress again that uncertainty in this
total of 894 counts does not affect our constraints on the \logn\
shape, as we show in \S~\ref{simulating}.

\im If the normalized flux distribution contains more than $N_{HST}=2184$ X-ray
sources between 0.05 and 30 counts (that is, more than the number of
X-ray-undetected \hst\ sources), we truncate the distribution at the
faint end and renormalize it so that
\begin{equation}
\sum_{i=1}^{n_{\rm flux}} N_{\rm src}(S_i)=N_{HST},
\end{equation}
while keeping $C_{\rm src}=894$ as per Eqn.~\ref{eqnnorm}.  This
effectively sets a minimum flux for the simulated X-ray sources. For
the power-law models described below, this is required for slopes
$\alpha \geq 1$.  The cutoff is always at $\leq0.3$ counts per 2 Ms
and does not significantly affect the final model fits (we might have
used a more sophisticated \logn\ distribution shape with a natural
low-flux cutoff, but it is not warranted by the data).  If the total
number of sources turns out to be less than $N_{HST}$, we do not
adjust the model flux distribution, since not all \hst\ sources are
necessarily X-ray sources.

\im We randomly assign fluxes to the 2184 \hst\ positions from the
distribution given by the normalized $N_{\rm src}(S)$.  For each
source, we use a model of the \chandra\ PSF (from the \chandra\ CALDB)
to create a $64\times64$ pixel X-ray probability map centered on the
\hst\ position, normalized to the model X-ray flux for that source and
$t_{\rm exp}$ at the source position.  The two-dimensional PSF model
averages over the 20 constituent CDF-N observations that have
different pointings (and thus off-axis angles) and exposure times.

\im We use the above model to create a simulated image of the sky.  We
produce Poisson realizations of the normalized background model
(described in \S\ \ref{meas}) and of the probability maps for the
sources, and add these realizations to create a simulated photon image.

In the process of creating this image, we eliminate ``detectable''
sources.  Before adding each source to the image, we derive the total
observed (post-Poisson scatter) source plus background counts $I$ in
the $r_{90}$ circle.  This corresponds to $I-S_B$ ``measured'' counts,
where $S_B$ is the value of the background model in the source region.
Based on the probability $f_c(I-S_B)$ that this number of observed
counts is detectable, we randomly assign the source as ``detected'' or
``undetected'', and do not add detected sources to the simulated image.

\end{enumerate}

\begin{figure}[t]
\epsscale{1.2}
\plotone{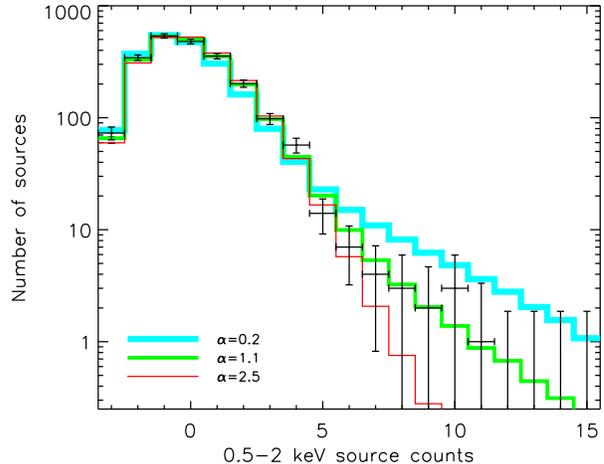}
\caption{Same as Fig.~\ref{figcounts}, with models for power-law
  \logn\ distributions.
\label{figlogn}}
\vskip0.5cm
\end{figure}

\begin{figure}[t]
\epsscale{1.2}
\plotone{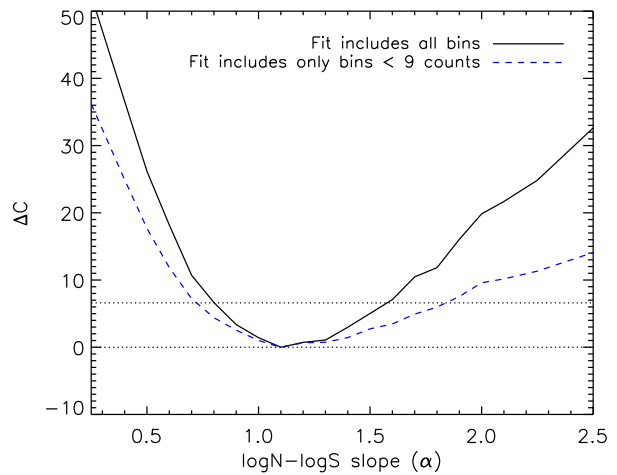}
\caption{
Variation in the $C$-statistic above the minimum ($\Delta C$) vs.  power-law slope
$\alpha$ of the source flux distribution.  The solid line shows the
results using all histogram bins, while the dashed line shows results
for only those bins with $< 9$ counts.
 The upper dotted line corresponds to the 99\% confidence limit ($\Delta C = 6.6$). 
\label{figlogn2}}
\vskip0.5cm
\end{figure}

We repeat steps 4 and 5 1000 times in order to average the random fluctuations
in each simulated image.  For each of these 1000 simulated sky images,
we derive the number of counts within $r_{90}$ for each of the \hst\
source regions, and produce a histogram of the background-subtracted
counts, exactly as for the sky image.  We average these 1000 histograms
to produce a predicted counts distribution for this \logn\ model and
compare it to the observed one.  

In comparing to the observed counts histogram, the $\chi^2$ test is
not appropriate because some of the bins have less than five sources,
so the uncertainties are not Gaussian.  Therefore, we calculate the
best-fit model and parameter confidence intervals using the
$C$-statistic \citep{cash79}, which is valid for fits to bins with low
numbers of counts and Poisson errors.  The probability distribution of
$\Delta C$, which is the difference between $C$ for the true value of
the free parameter and its measured minimum in $C$, is similar to the
$\chi^2$ distribution for 1 degree of freedom, which we verify for our
particular case using a simulation (described in \S~\ref{simulating}).
Thus, a 99\% confidence interval for one parameter corresponds to
$\Delta C\simeq 6.6$.  To calculate a rough goodness of fit (which is
not determined by the $C$-statistic), we also estimate $\chi^2$, using
errors (as shown in Fig.~\ref{figcounts}) given by
$\sigma_n=(n+0.75)^{1/2}+1$, which is an approximation of Poisson
uncertainties for $n$ sources per bin \citep{gehr86}.

\section{Results}

\label{results}

\subsection{Delta-function flux distribution}
As a simple but illuminating exercise, we begin with a ``delta
function'' flux distribution, such that all the X-ray-undetected \hst\
sources have either a single X-ray flux or zero flux [that is, $N_{\rm
src}(S_i)=0$ for all but one $S_i$].  We produce model distributions
for 0.5, 1, 1.5, 2, 3, and 4 photons per source per 2 Ms.  A sample
photon probability map and corresponding simulated image are shown in
Fig.~\ref{figmap}, and some of the resulting histograms are shown in
Fig.~\ref{figcounts}.  Because of Poisson fluctuations, even this
simple flux distribution produces a quite broad distribution in the
observed source counts.

Fig.~\ref{figcounts} shows that if all sources below the detection
threshold were relatively bright (with $S>3$ photons in 2 Ms), this
would produce too many \hst\ sources with 4--8 observed counts within
their $r_{90}$ source regions and too few with 1--2 counts.
Conversely, a distribution with only very faint sources ($S\leq1$ photon
in 2 Ms) gives too few sources with 7--8 counts in the source regions.
The data are fit well ($\chi^2=10$ for 18 degrees of freedom) if all
the undetected sources have fluxes of 1.5--3 photons per 2 Ms.

\subsection{Power-law flux distribution}
\label{powerlaw}
The actual flux distribution for the \hst\ sources is
likely not a delta function.  Here we consider the simple case of a
power-law distribution $N(>S)\propto S^{-\alpha}$, with $\alpha$
between 0.2 (flatter than the faint-end slope of the total CDF \logn)
and 2.5 (steeper than the observed \logn\ for galaxies; see \S\
\ref{intro}).  A sample photon probability map for a model with
$\alpha=1$ is shown in Fig.~\ref{figmap2}.

 The resulting counts histograms for these models are shown in
Fig.~\ref{figlogn}.  The $C$-statistic is minimized at $\alpha=1.1$,
for which the data are fit well ($\chi^2=8$ for 18 degrees of
freedom).  Notably, the negative bins in Figs.~\ref{figcounts} and
\ref{figlogn} are well-described by all models, indicating that the
statistical properties of the image are as expected and that we have
properly accounted for background.  The dependence of $\Delta C$ on $\alpha$ is shown by the solid line in
Fig.~\ref{figlogn2}.  At 99\% confidence ($\Delta C = 6.6$) we can
rule out flux distributions with $\alpha < 0.8$ and $\alpha > 1.6$.

\section{Verification}
\label{uncertainties}
\subsection{Background uncertainties}
\label{background}

\begin{figure}[t]
\epsscale{1.2}
\plotone{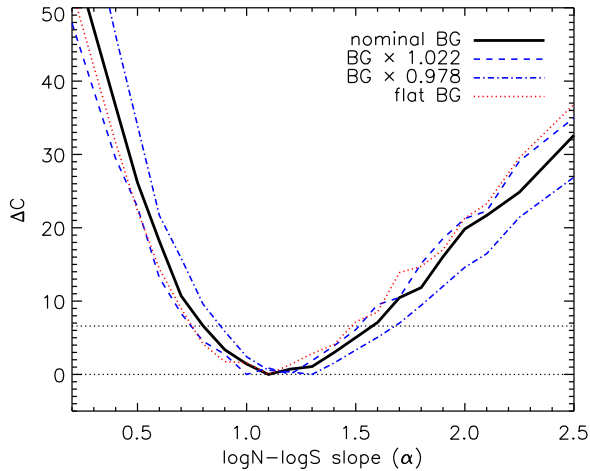}
\caption{Same as Fig.~\ref{figlogn2}, for different estimates of
  the diffuse background.  The thick line shows $\Delta C$ vs. $\alpha$ for the
  background as calculated in \S~\ref{meas} (as in Fig.~\ref{figlogn2}).  The blue
  dashed lines show results for the same spatial distribution of the
  background, scaled by 1.022 and 0.978.  The red dotted
  line shows results for a spatially uniform background, with a mean
  value as calculated in \S~\ref{meas}.  Varying the
  estimates of the background does not significantly change the
  constraints on $\alpha$. 
\label{figbg}}
\vskip0.3cm
\end{figure}

As described in \S~\ref{meas}, the number of background photons in the
\hst\ source regions is $\sim$4 times larger than the number of source
photons, so our subtraction of the background is a key component of
the analysis.  Because we measure the spatial distribution and
intensity of the background directly from the image itself, we expect
our background model to be sufficiently accurate.  Still, it is useful
to check that the results do not depend
strongly on our exact prescription for the background. 

First, we note that the data histogram shown in Figs.~\ref{figcounts}
and \ref{figlogn}, to which we perform the model fits, consists of
counts {\it after} subtraction of the background model.  We use
background-subtracted counts because they best represent the quantity
of interest, which is the distribution of photons associated with the
sources.  Since the same background model is subtracted from both the
data and the simulated images, we expect this procedure to have
minimal effect on the parameter constraints. To verify this, we also
calculate $\Delta C$ using the histogram of {\em total} observed
counts with no background subtraction, and find that the confidence
intervals are essentially identical (within $\approx5\%$) for the two
fits.

Next, we consider uncertainties in the (1) normalization and (2)
spatial distribution of the background model.  To address the impact
of a different background normalization, we repeat the analysis for
the \logn\ models described in \S~\ref{models}, but vary the overall
background level by $\pm$2.2\% (or 99\% confidence intervals for the
statistical error in the normalization given by Eqn.~\ref{eqnbgnorm}).
This in turn changes the total number of photons associated with the
\hst\ sources (which is 894 for the nominal background) by $\pm$10\%.

To address the effects of spatial variation, instead of using a
wavelet map of the background that approximates the chip gaps and
large-scale intensity variations, we repeat the analysis using a flat
(spatially uniform) background, normalized using Eqn.~\ref{eqnbgnorm},
which gives a background of 0.101 counts pixel$^{-1}$.  Each new background
model produces a slightly different histogram of background-subtracted
counts, in addition to new model distributions.

The values of $\Delta C$ versus $\alpha$ for these different
background models are shown in Fig.~\ref{figbg}.  The variation in the
background normalization changes the best-fit $\alpha$ by $<0.1$.
Because the background normalization uncertainty is statistical in
nature, we include this error in quadrature; it has only a small
effect on our final constraints on $\alpha$.  These alternative
backgrounds give similar goodness of fit ($6<\chi^2<11$) to the
nominal background.

\subsection{PSF uncertainties}
\label{psf}
We have also considered uncertainties in our model of the \chandra\
PSF, since scattered flux between source regions may affect the count
histograms.  To test this dependence, we repeat the calculations,
changing the width of the model PSF by $\pm 20\%$, or half the
variation in $r_{90}$ across our 2.2\arcmin\ region.  We expect that
the PSF model is significantly more accurate than $\pm 20\%$
\citep[e.g.,][]{alle03}, so this range is conservative.  Although we
change the PSF, we still use the $r_{90}$ circles as defined in
Eqn.~\ref{eqnrad} to derive the histogram of counts in the \hst\
regions.  Within this range of PSF widths, the flux scattered outside
the $r_{90}$ regions is between $\simeq$7\% and 13\%, which in turn
affects our estimate of the background normalization $N_B$ (as in
Eqn.~\ref{eqnbgnorm}) by a small amount ($\simeq$0.0015), as well as
our normalization of the model \logn\ (as in
Eqns.~\ref{eqnsninety}--\ref{eqnnorm}).

Properly accounting for these variations, we find that changing the
PSF width by $-20\%$ has a negligible effect, while changing the PSF
by $+20\%$ serves to increase the best-fit $\alpha$ by $\simeq0.2$.
The wider PSF causes additional flux overlap between the individual
sources, which produces more ``undetected'' sources with 8--15 counts
and more strongly rules out models with low $\alpha$.  However, even
for this conservative range of PSF widths, the uncertainty is still
significantly smaller than our statistical upper bound on $\alpha$.
We conclude that PSF uncertainties have no significant effect on our
constraints on the flux distribution.

There is a further slight uncertainty due to the fact that when we add
the PSF models to the simulated images, the source positions are
rounded to the nearest pixel, creating a positional offset up to
0.35\arcsec\ between the model PSF in the image and the center of the
source region.  To verify that this does not affect our results, we
repeated the analysis with the \hst\ source regions shifted to account
for this rounding offset; this produced negligible change in the
constraints on $\alpha$.

\subsection{Source detection completeness}
\label{complete}
We further consider uncertainty in the model of the source detection
completeness of the \alex\ catalog (given in \baue), which may affect
our modeling of the undetected \hst\ sources with $\gtrsim 8$ source counts.  For
example, some of our input \logn\ models are flatter, and have more
overlap between bright sources than the model used by \baue\ to
calculate the completness limits.  It is not immediately clear how
this source overlap would impact the source detection completeness.

One simple test is to examine the variation in $\Delta C$ if we only
include the bins at $\le 8$ counts, which are not strongly affected by
the completeness estimates.  This $\Delta C$ curve is shown by the
dashed line in Fig.~\ref{figlogn2}.  Although these fits give slightly
worse constraints on $\alpha$, they are consistent with the results
from the full histogram.  To more rigorously examine the impact of the
completeness, we repeat the full analysis using the simple case of a
hard flux limit; all sources with $>8$ source counts in the $r_{90}$
circle are detected, and all others are undetected (corresponding
roughly to the 9 count aperture-corrected limit for \alex). To perform
this comparison, we use only the histogram bins with $\leq8$ counts in
the fits.  This model gives nearly identical confidence limits (within
0.02 in $\alpha$) to the more sophisticated completeness model
described above. We conclude that uncertainty in the detection
completeness given in \baue\ should not significantly affect the
results.

\subsection{Verification of the fitting procedure}
\label{simulating}
Finally, we confirm the statistical validity of our constraints on
$\alpha$.  In particular, we verify that (1) $\Delta C$ is distributed
approximately like $\chi^2$ for 1 degree of freedom in our
particular case, and (2) our constraints are not affected by our
normalization of the model \logn\ distributions, which for
computational simplicity we set to match the observed 894 source
counts at the \hst\ source positions (thus ignoring the $\pm9\%$
$1\sigma$ statistical uncertainty in this value).

To this end, we perform a Monte Carlo simulation of the data and
our fitting procedure, with two simplifications to allow for fast
computations:  we do not consider source overlap (so that counts
observed at each source position are independent, and we do not need to
create simulated images), and  we do not account for the detection of
bright sources, only considering sources with $\leq10$ observed source
counts.  We define a model power-law \logn\ distribution with
our best-fit value of $\alpha=1.1$ and an average total flux of 1000 source photons observed
at 2500 source positions.  We also include a constant background flux
of 2 photons per source.

\begin{figure}[t]
\epsscale{1.2}
\plotone{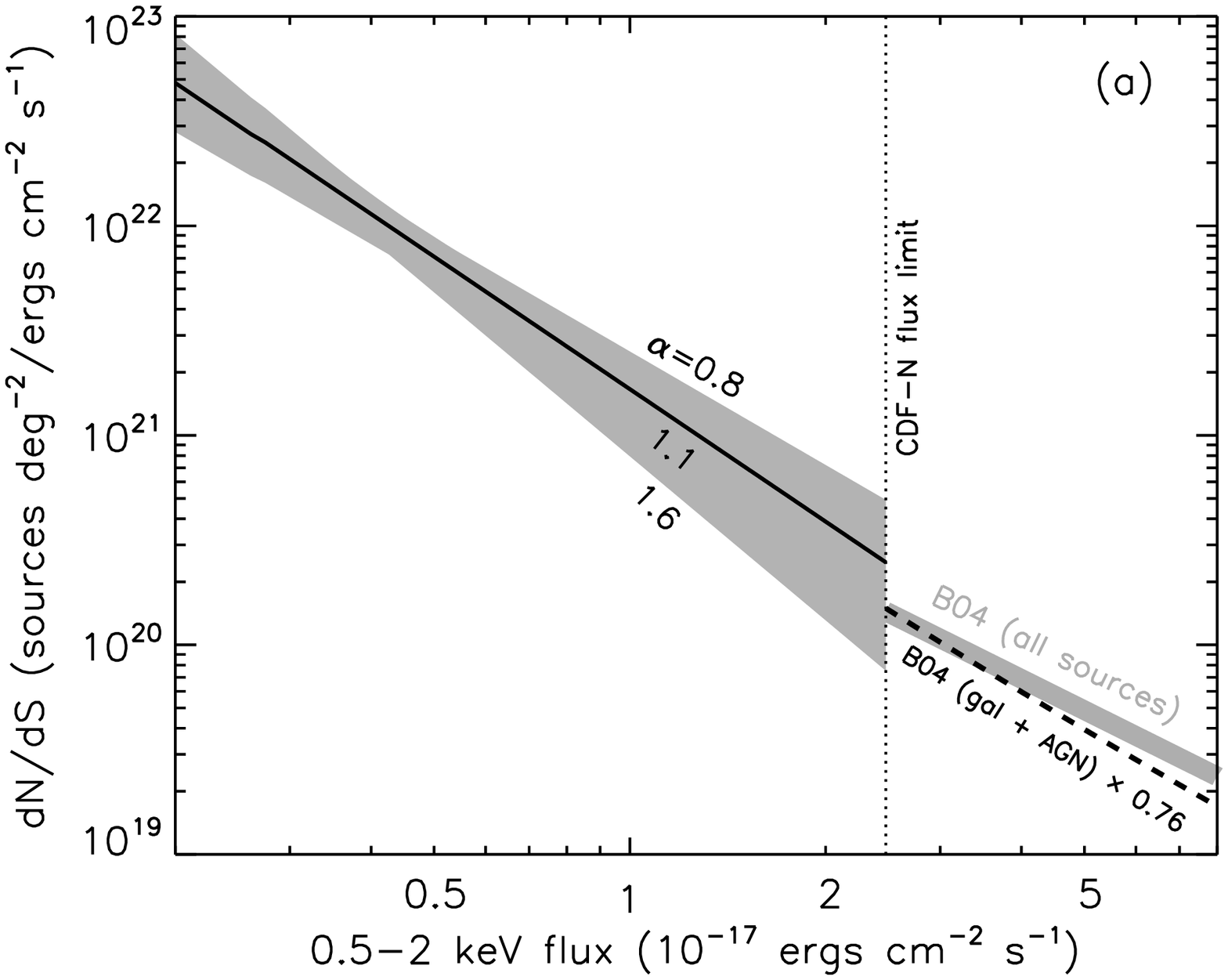}
\plotone{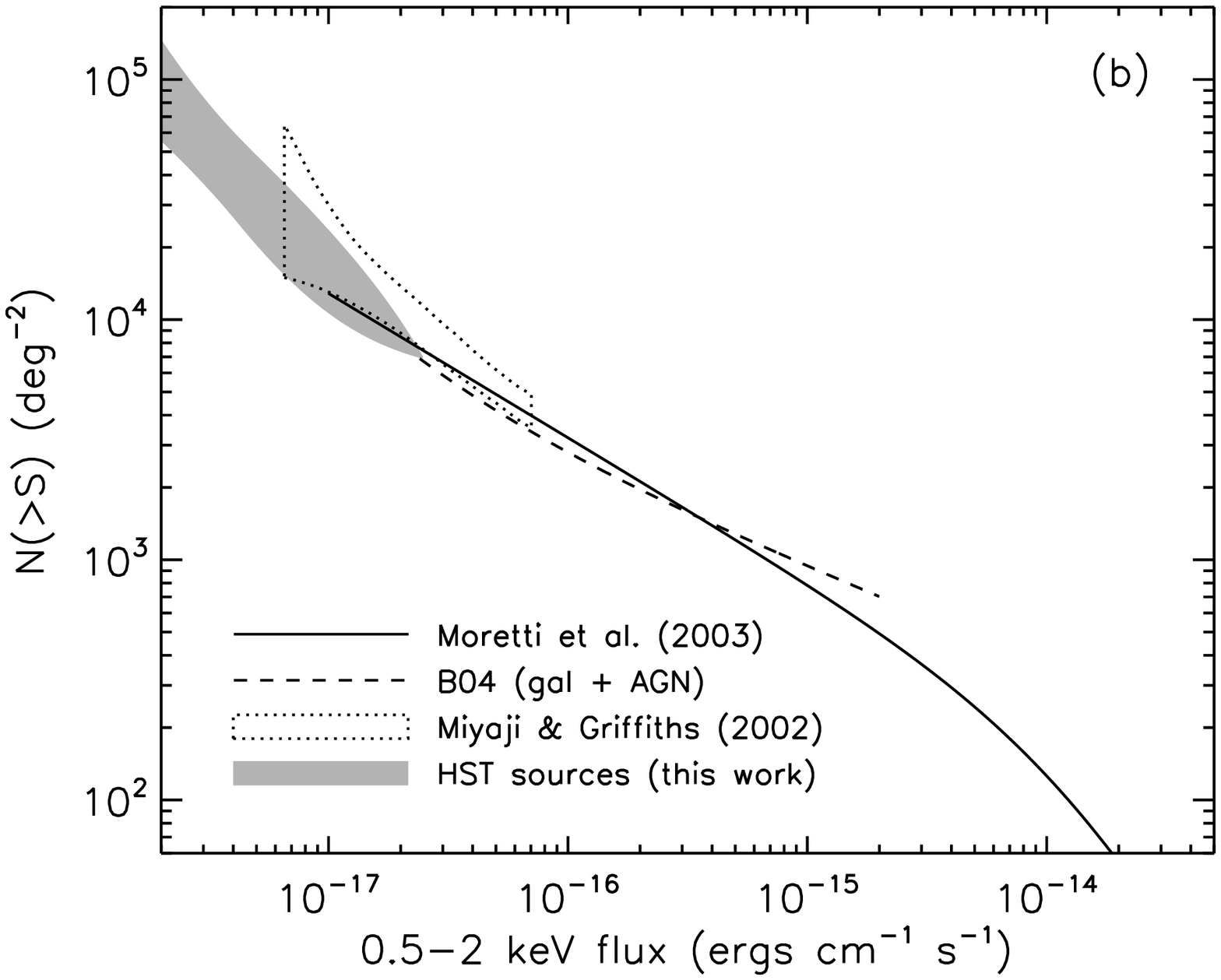}
\caption{ (a) The differential flux distribution (${\rm d}N/{\rm d}S$) for
 unresolved \hst\ sources, for the model power-law X-ray flux
 distributions allowed by the fits.  The shaded area shows the 99\%
 confidence intervals in power-law slope $\alpha$ and the
 normalization (see text for details).  For comparison, the lines at
 right show the best-fit \logn\ for all CDF-N sources (thick gray
 line) and AGNs plus galaxies (dashed line) from \baue.  For direct
 comparison with our model, the AGN plus galaxy curve is multiplied by
 0.76 (see text).  (b) The integral CDF \logn\ distribution, showing
 best-fit curves from \citet[][solid line]{more03} and the AGN plus
 galaxy fit from \baue\ (dashed line).  The gray area shows the 90\%
 confidence intervals from the fluctuation analysis on the 1 Ms CDF-N
 exposure by \citet{miya02}.  The hatched region shows the limits from
 the power-law fits to the unresolved \hst\ sources, with $N(>S)$ set
 to equal the \baue\ galaxy plus AGN fit at $S_{\rm 0.5-2\;
 keV}=2.4\times10^{-17}$ \flux.  Note that our constraints provide only lower
 limits to the total X-ray flux distribution, since some X-ray sources
 might not be detected by \hst.
\label{figlogn3}}
\vskip0.5cm
\end{figure}

We then create 1000 ``observed'' realizations of this flux
distribution, in which fluxes are assigned randomly to the 2500 source
positions.  For each source, the observed counts are drawn from a Poisson
distribution with a mean equal to the total source plus background flux.
We then follow the fitting procedure outlined in \S~\ref{models}, and fit each
realization of the data with model power-law \logn\ distributions
having slopes in the range $0.7<\alpha<1.6$.  For each realization, we
normalize the model \logn\ to produce, on average, the number
of total observed counts in that realization, rather than the input
average flux of 1000 source counts.  In this way, we take into account
statistical fluctuations in the total counts that we observe at the
\hst\ source positions.

For each realization, we calculate $C$ versus $\alpha$, and determine
the best-fit $\alpha$ from the minimum in $C$.  We find that in 99\%
of cases, the best-fit value of $\alpha$ lies $\Delta C<6.6$ from the
actual value of $\alpha=1.1$.  This confirms that (1) $\Delta C$ is
distributed very similarly to $\chi^2$ for 1 degree of
freedom, and (2) that statistical fluctuations in the total observed
counts do not affect our constraints on the slope $\alpha$.  These
fluctuations do, however, introduce uncertainty in the \logn\
normalization, which we include in our final results.

\vspace{1cm}

\section{Discussion}
\label{discussion}

\subsection{Constraints on the \logn\ distribution}

Our analysis shows that the distribution of photons in the CDF-N image
associated with X-ray-undetected \hst\ sources places significant
constraints on their underlying flux distribution.  Our final
constraints on the slope are $\alpha=1.1^{+0.5}_{-0.3}$.  The
corresponding normalization of the differential flux distribution is
$dN/dS=(1.3\pm0.3)\times10^{22}$ sources deg$^{-2}$ (\flux)$^{-1}$ at
$S_{\rm 0.5-2\; keV}=4\times10^{-18}$ \flux\ (all uncertainties are
99\% confidence).  In Fig.~\ref{figlogn3} (a), we show the $dN/dS$
distributions allowed by the fits.  The relatively small uncertainty
at $S_{\rm 0.5-2\; keV}\sim4\times10^{-18}$ \flux\ is a result of the
fact that we restrict our \logn\ models to be power-law in shape.

Our best-fit power-law model ($\alpha=1.1$)
connects well with that of the observed CDF-N \logn.  In
Fig.~\ref{figlogn3} (a) we compare it to the \baue\ fits for (1)
the full CDF \logn, which has $\alpha\simeq0.7$, and (2) the sum of
the galaxies plus AGNs power laws fitted separately, which near the
CDF-N flux limit can be approximated by a single power law with
$\alpha\sim 1$.  While these \logn\ curves represent all X-ray
sources, our fits constrain the flux distributions for \hst\ sources
only.  Therefore, to account for the fact that only 16 out of the 21
detected X-ray sources in our 2.2\arcmin\ circle have \hst\
counterparts, we show the \baue\ galaxy plus AGN curve multiplied by
this fraction (0.76).  At the CDF-N limit, the normalizations of both
\baue\ curves are consistent with our allowed model distributions, but
the slope of the AGN plus galaxy \logn\ more closely matches our
best-fit model for the fainter \hst\ sources.  This indicates that the
X-ray-unresolved \hst\ sources might represent the faint end of the
resolved sub-population of normal and starburst galaxies, and that
deeper \chandra\ observations might resolve significant numbers of
these sources.  

Fig.~\ref{figlogn3} (b) shows fits to the integrated \logn\ in the
 CDFs.  Shown are with the best-fit curve from \citet{more03}, as well
 as the AGN plus galaxy fit from \baue\ (which only included sources
 fainter than $S_{\rm 0.5-2\; keV}=2\times10^{-16}$ \flux).  We also
 show the constraints from our fits, by setting $N(>S)$ equal the
 \baue\ galaxy plus AGN curve at $S_{\rm 0.5-2\; keV}=2.4\times10^{-17}$
 \flux\ (this is necessary because we analyze only the {\em undetected}
 sources).  The unresolved \logn\ distributions allowed by our fits
 are consistent with the constraints obtained by the fluctuation
 analysis on the first 1 Ms of data by \citet{miya02}, and extend to lower fluxes.

\begin{figure}[t]
\epsscale{1.2} \plotone{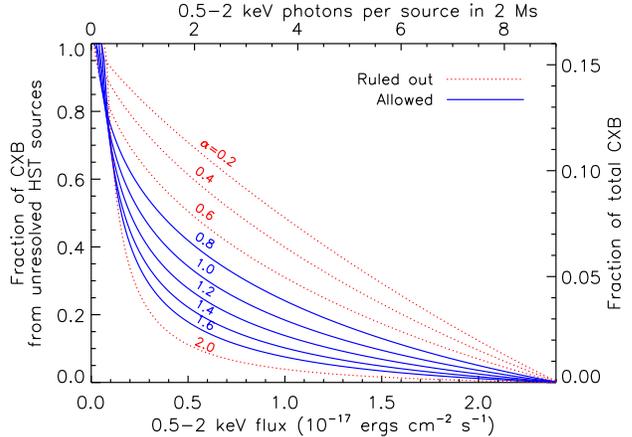}
\caption{The fraction of the unresolved extragalactic 0.5--2 keV CXB
from undetected \hst\ sources that can be resolved as a function of
limiting \chandra\ flux, for power-law model \logn\ distributions.
The current limit for 2 Ms is at the end of the $x$-axis.  Models with
$\alpha<0.8$ and $\alpha>1.6$ are ruled out by the model fits (see Figs. \ref{figlogn2} and \ref{figbg}).
The left $y$ axis represents the flux associated with X-ray-undetected
\hst\ sources only (16\% of the total soft CXB, \hmn).  The right $y$ axis
gives the corresponding fraction of the total CXB.
\label{figfrac}}
\vskip0.5cm
\end{figure}

\subsection{Prospects for deeper \chandra\ exposures}

How much of the remaining extragalactic CXB would be resolved by
deeper \chandra\ exposures depends largely on the shape of the flux
distribution for unresolved sources.  For a shallow \logn\ curve (i.e., low $\alpha$), most
of the unresolved CXB would come from relatively bright sources that
could be detected with moderately deeper observations.  For a steeper
\logn, however, the bulk of the flux comes from much
fainter sources.

As measured by \hmn, 70\% of the unresolved (or 16\% of the total)
1--2 keV CXB is associated with \hst\ sources.  Fig.~\ref{figfrac}
shows the fraction of this flux that would be resolved as a function
of survey depth, for the power-law \logn\ models used in
\S~\ref{analysis}.  Here we consider the CXB to be ``resolved'' down
to the flux of the faintest detected sources, ignoring
incompleteness in the source detection (e.g., the \alex\ catalog is
only $\sim$10\% complete at its nominal 9 count detection limit).
The models with high and low $\alpha$ (Fig.~\ref{figfrac}, dotted lines) are
ruled out by our present analysis.

Let us consider further observations that are 5 times more sensitive
than the current CDF-N, with a 0.5--2 keV flux limit of
$5\times10^{-18}$ \ergs.  Source detection in such observations would
likely become background-limited, so this sensitivity would require
exposures more than 5 times the existing one.  For these exposures,
the range of $\alpha$ allowed by the fits ($0.8<\alpha<1.6$)
corresponds to resolving between $\sim$20\% and 40\% of 0.5--2 keV
flux from the undetected \hst\ sources.  This would make up 13\%--30\%
of the unresolved (or 3\%--7\% of the total) soft CXB.  For the best-fit
power law model with $\alpha=1.1$, $\sim$30\% of the flux from the
undetected \hst\ sources would be resolved.

Of course, the present analysis does not constrain the fluxes from
those undetected X-ray sources that do not have \hst\ counterparts.
Such sources can contribute up to an additional $\simeq$30\% of the
unresolved (or $\simeq$6\% of the total) soft CXB.  In the most
optimistic scenario, all these sources would be detected in deep
exposures, and the \hst\ sources would have the flattest allowed
\logn\ slope.  In this optimistic case, \chandra\ exposures 5 times
more sensitive than the existing observations (i.e., $>10$ Ms) would resolve up to
$\sim$60\% of the remaining extragalactic soft CXB.

\acknowledgements 

We thank A.~Vikhlinin, C.\ Jones, R.\ Narayan, K.\ Lai, M.\ Weisskopf,
and S.\ O'Dell for fruitful discussions and comments, and the referee
for useful suggestions. R.C.H. was supported by a NASA GSRP Fellowship
and a Harvard Merit Fellowship, and M.M. by NASA contract NAS8-39073 and
\chandra\ grant G06-7126X.

\clearpage

\end{document}